\documentclass[conference,10pt]{IEEEtran}
\IEEEoverridecommandlockouts
\usepackage[letterpaper,
            left=1in,right=1in,top=1.91cm,bottom=2.55cm]{geometry}

\usepackage[font=footnotesize,skip=5pt]{caption}
\usepackage{lipsum}
\usepackage{cite}
\usepackage[T1]{fontenc}
\usepackage[cmintegrals]{newtxmath}
\usepackage{array}
\usepackage[caption=false,font=footnotesize]{subfig}
\usepackage{url}

\usepackage{balance}
\usepackage{graphicx}
\usepackage{algorithm}
\usepackage{algpseudocode}
\usepackage{pifont}
\usepackage{amsmath}
\usepackage{float}
\usepackage{multicol}
\usepackage{mwe}
\usepackage{epstopdf}
\usepackage[utf8]{inputenc}
\DeclareGraphicsExtensions{.eps}
\usepackage{array}
\usepackage{amsmath}
\usepackage{mathtools}
\usepackage[subnum]{cases}
\usepackage{color,soul}% for highlights

\makeatother
\begin{document}

\title{User Association and Bandwidth Allocation for Terrestrial and Aerial Base Stations\\ with Backhaul Considerations }

\author{\IEEEauthorblockN{Elham Kalantari\IEEEauthorrefmark{1},
		Irem Bor-Yaliniz\IEEEauthorrefmark{2},
		Abbas Yongacoglu\IEEEauthorrefmark{1}, and 
		Halim Yanikomeroglu\IEEEauthorrefmark{2}\thanks{This work is supported in part by Huawei Canada Co., Ltd.}}
	\IEEEauthorblockA{\IEEEauthorrefmark{1}School of Electrical Engineering and Computer Science\\
		University of Ottawa, Ottawa, ON, Canada, Email: \{ekala011, yongac\}@uottawa.ca}
	\IEEEauthorblockA{\IEEEauthorrefmark{2}Department of Systems and Computer Engineering\\
	Carleton University, Ottawa, ON, Canada, Email: \{irembor, halim\}@sce.carleton.ca}}

\maketitle

\begin{abstract}
Drone base stations (DBSs) can enhance network coverage and area capacity by moving supply towards demand when required. This degree of freedom could be especially useful for future applications with extreme demands, such as ultra reliable and low latency communications (uRLLC). However, deployment of DBSs can face several challenges. One issue is finding the 3D placement of such BSs to satisfy dynamic requirements of the system. Second, the availability of reliable wireless backhaul links and the related resource allocation are principal issues that should be considered. Finally, association of the users with BSs becomes an involved problem due to mobility of DBSs. In this paper, we consider a macro-BS (MBS) and several DBSs that rely on the wireless links to the MBS for backhauling. Considering regular and uRLLC users, we propose an algorithm to find efficient 3D locations of DBSs in addition to the user-BS associations and wireless backhaul bandwidth allocations to maximize the sum logarithmic rate of the users. To this end, a decomposition method is employed to first find the user-BS association and bandwidth allocations. Then DBS locations are updated using a heuristic particle swarm optimization algorithm. Simulation results show the effectiveness of the proposed method and provide useful insights on the effects of traffic distributions and antenna beamwidth.  
\end{abstract}

\IEEEpeerreviewmaketitle
%\vspace{-0.01cm}
\section{Introduction}
For future wireless networks, three main use cases are being considered: Enhanced mobile broadband (eMBB), massive machine-type communications (mMTC), and ultra reliable and low latency communications (uRLLC)~\cite{osseiran2014scenarios}. All of these use cases have challenging demands and they are very different from each other. For instance, while mMTC applications tolerate low data rates and large delays, uRLLC applications can be very demanding to provide reliability and low-latency requirements. In such cases, providing isolated routes and caching to reduce latency, and allocating more wireless resources to provide reliability may be necessary~\cite{urllc}.

To increase the agility and flexibility of the network, drones can be integrated into the wireless network as flying base stations (BSs). It is a promising approach which assists the ground network of BSs by temporarily increasing capacity and/or coverage whenever and wherever it is required, especially when the demand occurs in a rather difficult-to-predict manner \cite{eliVTC}. However, utilizing drone-BSs (DBSs) has many challenges as well~\cite{iremmag}, such as their positioning, wireless backhauling, resource allocation for access and backhaul links, and user association, all of which are considered in this study. 

Prior works in this area investigate the integration of flying platforms such as drones in cellular networks with respect to their placement, various use cases, and design and management issues. In \cite{iremmag}, multi-tier drone networks were introduced to complement terrestrial heterogeneous networks (HetNets) and the advancements and challenges related to the operation and management of DBSs were investigated. In \cite {mohammadmag,Awais1,Awais2}, a network of flying platforms was considered as backhaul/fronthaul hubs for small cells via free-space-optics /mmWave links. Finding the efficient DBS placement under different assumptions and considerations were studied in \cite{eliVTC,eliICC,iremconf, AlzenadLetter, MozaffariSBD15a}. In \cite{eliVTC}, the minimum number of DBSs and their 3D placements were found using a heuristic algorithm in order to serve high data rate users. It was observed that altitude is an important factor in DBS deployment to tackle coverage or capacity issues. In \cite{eliICC}, the optimal 3D backhaul-aware placement of a DBS in 2 different approaches, namely network-centric and user-centric, was found and the robustness of the network with respect to the users' displacement was examined. In \cite{iremconf}, the 3D placement of a DBS to maximize the number of covered users through numerical methods was found. In \cite {AlzenadLetter}, an algorithm was proposed to find the maximum number of covered users while the transmit power is minimized by decoupling the problem to vertical and horizontal dimensions.

In this study, uRLLC users with delay-sensitive applications co-exist with regular eMBB users. To the best of our knowledge, this is the first time in the literature that backhaul resource allocation, user association considering user types, and 3D placement of DBSs are considered jointly. Moreover, on the contrary to many other studies mentioned above, the existence of MBSs is also taken into account. A novel problem formulation considering fairness and a numerical solution method is provided.  

\begin{figure}[t]
	\begin{center}
		\includegraphics[width=3in]{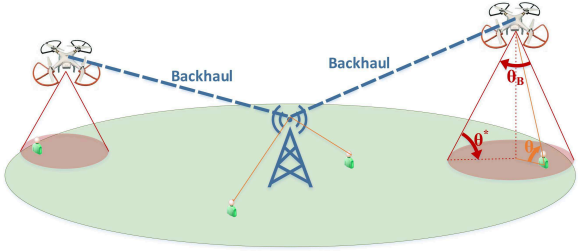}\\
	\end{center}
	\caption{Graphical illustration for integration of DBSs in a cellular network.}
	\label{illustrative}
\end{figure}

The rest of this paper is organized as follows. In Section II, the system model along with the problem formulation is presented. The algorithm is proposed in Section III, followed by the performance evaluations in Section IV. Finally, conclusions are drawn in Section V.

\section{System Model}
We consider a downlink wireless HetNet including two tiers of BSs, an MBS and a number of DBSs. DBSs are utilized to serve as small-cells assisting the wireless network in cases where the existing infrastructure is insufficient to address the demand. They are assumed to utilize wireless connection for both access and backhaul links. On the one hand, wireless links provide a mobility advantage to DBSs such that they can be positioned with respect to the users, which can increase spectral efficiency and decrease average path loss. On the other hand, wireless links can be less reliable compared to wired connections, and energy expenditure increases too. Therefore, a careful system design is key. 

We denote by $\mathcal{I}$ the set of users, and $\mathcal{J}$ the set of BSs. We use $i\in \mathcal{I}=\{1,2,...,N\}$ and $j\in \mathcal{J}=\{0,1,...,M\}$ to index users and BSs, respectively. Index $0$ in $\mathcal{J}$ denotes the only MBS considered in this system. We assume that high capacity fiber links carry information from the MBS to the core network; therefore, there is no congestion in the backhaul link of the MBS. We also assume that in-band wireless backhaul is employed for DBSs and the MBS is utilized as a hub to connect DBSs to the network. To avoid self-interference, orthogonal frequency channels in the backhaul and access side of the DBSs is employed. Therefore, part of the bandwidth is shared between the access side of the MBS and DBSs and the remainder is dedicated for the backhaul of DBSs. The free space path loss (FSPL) according to the Friis equation is considered for backhaul links. The FSPL is given as $20 \log(\frac{4\pi f_c d}{c})$, where $f_c$ is the carrier frequency, $c$ stands for the speed of light, and $d$ stands for the distance between the transmitter and receiver.

We assume there are wireless point-to-point $X_n$ links between BSs, which do not interfere with access and backhaul links. Considering non-ideal $X_n$ connections and the energy cost of wireless links, real-time coordination for interference management among BSs may not be efficient. Hence, to decrease inter-cell interference, reverse time division duplex is employed, which uses reversed uplink/downlink time slot configurations for MBS and DBSs \cite{7386685}. When the MBS is in downlink mode, the DBSs are in the uplink mode. As a result, the only interference the MBS users receive is from the DBS users, which is negligible as the transmit power of user equipments is lower than that of an MBS.  

\subsection{Channel Models}
The air-to-ground path loss depends on the height of a DBS and the elevation angle between a DBS and a ground user, denoted by $\theta$ in Fig.~\ref{illustrative}. There are mainly two propagation groups, corresponding to the receivers with line-of-sight (LoS) connections and those with non-line-of-sight (NLoS) connections which can still receive the signal from the transmitter due to strong reflections and diffractions \cite{7037248}. The total power reduction of a signal transmitted from a DBS to a ground user can be written in decibel form as
\begin{equation}\label{fspl}
\textsf{PL (\textrm{dB})} = \textsf{FSPL} +\psi_i,
\end{equation}
where $\psi_i, i=\{\textrm{LoS},\textrm{NLoS}\}$ shows the excessive path loss due to the LoS or NLoS channel between the DBS and the user. A Gaussian distribution can be used to model $\psi_i$ as  $\mathcal{N}(\mu_i,\sigma_i)$, where $\mu_i$ is the mean excessive path loss shown by a constant value depending only on the environment, and $\sigma_i = k_i\exp (-l_i\cdot\theta)$, where $k_i$ and $l_i$ are frequency- and environment-dependent parameters. 

The probability of having a LoS connection between a DBS and a user can be formulated as \cite{6863654}
\begin{equation}\label{LoS}
P(\textrm{LoS}) = \frac{1}{1+a \exp (-b(\theta -a))},
\end{equation}
where $a$ and $b$ are constant values depending on the environment and $\theta$ is equal to $\frac{180}{\pi}\arctan (\frac{\rho}{\delta})$, where $\rho$ and $\delta$ are the altitude of the DBS and the projection of its distance from the user on the ground, respectively.

The antenna gain can be approximated by \cite{7486987}
\begin{equation}
G = 
\begin{dcases}
G_0, -\frac{\theta_B}{2}\le \phi \le \frac{\theta_B}{2},\\
g(\phi), \text{otherwise},
\end{dcases}
\end{equation}
where $|\phi|=90 - \theta$, $\theta_B$ denotes the DBS directional antenna's half-power beamwidth and $G_0  \approx \frac{30,000}{\theta_B^2}$ is the maximum gain of the directional antenna \cite{Balanis}. We assume $g(\phi)$ is negligible.

We adopt the MBS channel model from 3GPP TR 36.942 \cite{3GPP}. The average path loss in dB can be expressed as $128.1 +37.6 \log_{10}(d')$, where $d'$ is the distance between the transmitter and receiver in kilometers. Also, the lognormal shadowing with standard deviation 10 dB is assumed. Moreover, an omni-directional antenna is considered in our model. %with the antenna gain of 15 dBi 

\subsection{Problem Formulation}
The mobility of the DBSs and different types of users require that the following key issues are considered to provide wireless services efficiently:
\begin{itemize}
\item Finding the locations of DBSs,
\item Determining the user-BS associations with consideration to user type,
\item Bandwidth allocation for access and backhaul links.
\end{itemize}
A user cannot be associated with more than one BS; therefore,
\begin{equation}
\sum_{j\in \mathcal{J}}x_{ij} = 1 ,\forall{i\in \mathcal{I}},
\end{equation}
where $x_{ij} \in \{0,1\}$ is the binary association indicator variable for user $i$ and BS $j$, and 1 indicates association.

If the total bandwidth in the network is unity, we can denote the part assigned to backhaul of DBSs with $\alpha$, and the part assigned to the access of both the MBS and DBSs with $1-\alpha$. The total amount of resources allocated by each BS to all the users cannot exceed its available bandwidth; therefore,
\begin{equation}
\sum_{i \in \mathcal{I}} x_{ij} \cdot y_{ij} \le 1-\alpha, \forall j \in \mathcal{J},
\end{equation}
where $y_{ij} \in [0,1]$ is resource amount that is assigned to user $i$ from BS $j$.

The total data rate a DBS can support should not exceed its backhaul capacity; so,
\begin{equation}
\sum_{i \in \mathcal{I}} R_{ij} \le C_j, \forall{j \in \mathcal{J}\backslash{0}}, \label{BH1}
\end{equation}
where $R_{ij}$ is the total data rate of user $i$ receiving from BS $j$ and $C_j$  is the backhaul capacity of DBS $j$. Assuming Shannon capacity is achieved, $C_j$ can be written as
\begin{equation}
C_{j} = \alpha \cdot r_{j0},
\end{equation} 
where $r_{j0}=\log_2(1+\gamma_{j0})$ and $\gamma_{j0}$ is the received signal-to-noise ratio (SNR) at the DBS $j$ from the MBS for the backhaul connection. It is equal to $\gamma_{j0} = \frac{P_0 h_{j0}}{\sigma^2}, j\in \mathcal{J}\backslash 0$, where $P_0$ denotes the transmit power of the MBS, $h_{j0}$ stands for the channel gain between the MBS and DBS $j$, and $\sigma^2$ denotes the noise power level. Similarly, 
\begin{equation}
R_{ij} = x_{ij} \cdot y_{ij} \cdot r_{ij},
\end{equation}
where $r_{ij}$ is the instantaneous achievable rate of user $i$ associated with BS $j$, and 
\begin{equation}\label{rate}
r_{ij} = \log_2 (1+\gamma_{ij}). 
\end{equation}
Then, 
\begin{numcases}
{\gamma_{ij} =}
\frac{P_{j} h_{ij}}{\sum_{l\in \mathcal{J},l\ne j} P_{l} h_{il}+\sigma^2},j\in \mathcal{J}\backslash 0, \label{SINR}\\
\frac{P_{j} h_{ij}}{\sigma^2},j= 0, \label{SNR}
\end{numcases}
%\begin{subequations}
%\begin{eqnarray}
% eq \\
% eq \\
% eq
%\end{eqnarray}
%\end{subequations}
where $P_j$ denotes the transmit power of BS $j$ and $h_{ij}$ stands for the channel gain between user $i$ and BS $j$. Note that, \eqref{SINR} is the received signal-to-interference-plus-noise ratio (SINR) of user $i$ if it is associated with DBS $j$,  while \eqref{SNR} is the received SNR of user $i$ if it is associated with the MBS.

As wireless backhaul links for DBSs may increase the latency, we assume that uRLLC users utilize a delay-sensitive application, and can only be associated with the MBS, which can be formulated as  
\begin{equation}
\sum_{j\in \mathcal{J}\backslash 0}x_{ij} \le 1-\tau_{i} ,\forall{i\in \mathcal{I}},
\end{equation}
where $\tau_i \in \{0,1\}$; $\tau_i = 1$ indicates that the user $i$ is delay-sensitive and $\tau_i = 0$ indicates the opposite. 

The probability of LoS connection is usually high in DBSs and as all of them share the same bandwidth, it might cause high interference. To mitigate such interference, we assume that DBSs are equipped with directional antennas and that only the users in the footprint of a DBS antenna coverage can be served by it. It can be formulated as 
\begin{equation}
x_{ij}\cdot (\theta_{ij}-\theta^*) \ge 0 ,\forall{i\in \mathcal{I}},\forall{j\in \mathcal{J}\backslash 0},
\end{equation}
where $\theta_{ij}$ is the elevation angle between user $i$ and DBS $j$, and $\theta^* = 90 - \frac{\theta_B}{2}$.

\begin{figure}[t]
	\begin{center}
		\includegraphics[width=2.9in]{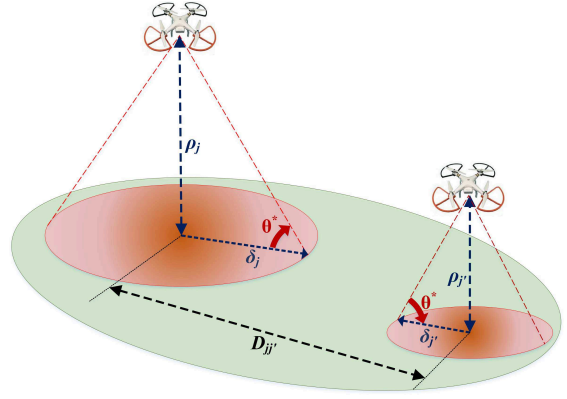}\\
	\end{center}
	\caption{Distance between DBSs, $D_{jj'}$, must be large enough to prevent interference, as derived in~\eqref{min-distance}.}
	\label{distance}
\end{figure}

Note that the coverage radius, $\delta_j$, of DBS $j$ is related to its altitude as $\tan (\theta^*) = \frac{\rho_j}{\delta_j}$, where $\rho_j$ is the altitude of DBS $j$. Accordingly, the minimum distance required between two DBSs to prevent interference can be written as follows (Fig.~\ref{distance}):
\begin{equation}
\frac{\rho_j+\rho_{j'}}{\tan(\theta^*)}, \forall j, j' \in \mathcal{J}\backslash 0.
\end{equation}

In order to prevent over-loading some DBSs (e.g., if the users are clustered) we consider fairness. Therefore, a logarithmic utility function is assumed, where $U(R_i) = \log R_i$. Hence, the problem formulation also considers fairness and can be cast as the following mixed-integer optimization problem:
\begin{equation}
\max_{ \{{l_{j\in \mathcal{J}\backslash 0}} \} ,\{ x_{ij} \} , \{ y_{ij} \}, \{ \alpha \} } \sum_{i \in \mathcal{I}} U(R_i) 
\end{equation}
subject to:
\begin{eqnarray}
\sum_{j\in \mathcal{J}}x_{ij} = 1 ,\forall{i\in \mathcal{I}}, \label{user-association}\\
\sum_{i \in \mathcal{I}} x_{ij} \cdot y_{ij} \le 1-\alpha, \forall j \in \mathcal{J},\\
\sum_{i\in \mathcal{I}}x_{ij} \cdot y_{ij} \cdot r_{ij}  \le \alpha \cdot r_{j0}, \forall j \in \mathcal{J}\backslash 0,\\
\sum_{j\in \mathcal{J}\backslash 0}x_{ij} \le 1-\tau_{i} ,\forall{i\in \mathcal{I}},\label{delay}\\
x_{ij}\cdot (\theta_{ij}-\theta^*) \ge 0 ,\forall{i\in \mathcal{I}},\forall{j\in \mathcal{J}\backslash 0},\label{antenna}\\ 
D_{jj'} \ge \frac{\rho_j+\rho_{j'}}{\tan(\theta^*)}, \forall j, j' \in \mathcal{J}\backslash 0 , j \ne j', \label{min-distance}\\
x_{ij} \in \{0,1\},\label{user-indicator}\\
\alpha \in [0,1]\label{alpha},\\
y_{ij} \in [0,1-\alpha],
\end{eqnarray}
where $l_{j\in \mathcal{J}\backslash 0}$ is the 3D location of DBS $j$ and $R_i = \sum_{j \in \mathcal{J}} R_{ij}$ is the total rate of user $i$.

Equal resource allocation is the optimal allocation for the logarithmic utility \cite{6497017}; therefore, $y_{ij}=\frac{1-\alpha}{\sum_{k \in \mathcal{I}}x_{kj}}$ and the problem is transformed to
\begin{equation}
\max_{ \{{l_{j\in \mathcal{J}\backslash 0}} \} ,\{ x_{ij} \} , \{ \alpha \} } \sum_{i \in \mathcal{I}} \sum_{j \in \mathcal{J}}x_{ij}  \log \frac{r_{ij}\cdot (1-\alpha)}{\sum_{k \in \mathcal{I}}x_{kj}}
\end{equation}
subject to:
\begin{eqnarray}
\sum_{i\in \mathcal{I}}x_{ij} \cdot\frac{1-\alpha}{\sum_{k \in \mathcal{I}}x_{kj}} \cdot r_{ij}  \le \alpha \cdot r_{j0}, \forall j \in \mathcal{J}\backslash 0 \label{BH2},
\end{eqnarray}
{\centering (\ref{user-association}), and (\ref{delay}) - (\ref{alpha}).\par}
  
Even after the above simplification, the optimization problem has a non-convex objective function with non-linear constraints with a combination of binary and continuous variables. In other words, it is a non-convex mixed-integer, NP-hard optimization problem. 

\section{Proposed Algorithm}
To alleviate the difficulties mentioned in the preceding section, we first relax the binary cell association indicator, $x_{ij}$. It upper bounds the performance and corresponds to the case where users can be associated to multiple BSs. Then, for fixed DBS locations, the optimization problem becomes a separable problem in $x_{ij}$ and $\alpha$ and can be solved through a primal decomposition algorithm \cite{7386685}. This procedure includes three main processes:
\begin{enumerate}
\item The user-BS association problem can be written as a convex subproblem for a fixed $\alpha$, as 
\begin{equation}
\max_{ \{ x_{ij} \} }  \sum_{i \in \mathcal{I}} \sum_{j \in \mathcal{J}}x_{ij}  \log \frac{r_{ij}\cdot (1-\alpha)}{\sum_{k \in \mathcal{I}}x_{kj}}
\end{equation}
subject to:
{\centering (\ref{user-association}), (\ref{delay}), (\ref{antenna}), (\ref{user-indicator}), and (\ref{BH2}).\par}
This can be solved with convex optimization tools efficiently.

\item After finding $x_{ij}$ and rounding it, the following master problem, which is also convex, is solved.
\begin{equation}
\max_{\alpha } \log (1-\alpha)
\end{equation}
subject to:
{\centering (\ref{alpha}) and (\ref{BH2}).\par}
Each iteration of the master problem requires solving the subproblem and updating $x_{ij}$ variables.

\item After finding variables $x_{ij}$ and $\alpha$, the location of DBSs is updated through the particle swarm optimization (PSO) method by maximizing the utility function~\eqref{PSO}. The required constraints are added as penalty functions to this utility function.

% The algorithm starts with a population of random solutions and iteratively tries to improve the candidate solutions with regards to a given measure of quality. The best experience of each candidate as well as the best global experience of all the candidates in all iterations are recorded and the next movement of the candidates is influenced by these items \cite{488968}. The required constraints are added as penalty functions to the utility function in~\eqref{PSO}.

\end{enumerate}
Processes 1-3 are repeated until convergence is reached. The proposed algorithm is summarized in Algorithm \ref{Alg1}.
\begin{figure*}[t]
\begin{equation}
\small
\sum_{i \in \mathcal{I}} \sum_{j \in \mathcal{J}}x_{ij}  \log \frac{r_{ij}\cdot (1-\alpha)}{\sum_{k \in \mathcal{I}}x_{kj}} -
\sum_{j, j'\in \mathcal{J} \backslash 0 , j \neq j'} \Big( \tan (\theta^*) \cdot D_{jj'} - \rho_j - \rho_{j'}\Big) -
\sum_{j \in \mathcal {J}\backslash 0} \Big(\alpha \cdot r_{j0} -\sum_{i\in \mathcal{I}}x_{ij} \cdot\frac{(1-\alpha) \cdot r_{ij}}{\sum_{k \in \mathcal{I}}x_{kj}}   -
 \sum_{i \in \mathcal{I}} x_{ij} \cdot (\theta_{ij}-\theta^*) \Big )
\label{PSO}
\end{equation}
\hrule
\end{figure*}

\begin{algorithm}[tb]
\small
\caption{Finds 3D locations of DBSs, user-BS association and bandwidth allocation for access and backhaul of DBSs.}
\label{Alg1}
\begin{algorithmic}[1]
\State \textbf{Inputs:} Users' locations, number of DBSs.
\State \textbf{Initialization:} Cluster the users based on the number of DBSs using k-means clustering. Assume the initial location of DBSs is the center of the clusters. Set $t=1$, $m(t)= U(t)-U(t-1)$, $t'=1$, $n(t')=U(t')-U(t'-1)$; define $\alpha(1)=a$, $m(1)=n(1)=M$, where $M$ is a big number. $\nu$ and $\epsilon$ are small positive numbers. 
\While  {$n(t') \ge \nu$ }
\While{$m(t) \ge \epsilon$ }
\State Find $x_{ij}(t), U(t)$. Round $x_{ij}(t)$.
\State $t = t+1$.
\State Find $\alpha(t), U(t)$.
\EndWhile
\State Find $U(t')$.
\State $t' = t' +1$.
\State Update the 3D locations of DBSs using PSO algorithm. Find $U(t')$.
\EndWhile
\end{algorithmic}
\end{algorithm} 

\section{Performance Evaluation}
We consider an urban region with total area 250000 m$^2$, which is served by one MBS in the center of the area, and 3 DBSs at locations and altitudes to be determined. 
We assume that users have a $Mat\acute{e}rn$ distribution, which is a doubly Poisson cluster process \cite{martin-haenggi}. The heterogeneity of the users distribution is measured by the coefficient of variation (CoV) of the Voronoi area of the users \cite{7110500,7425184}. It is defined as $\frac{1}{0.529} \frac{\sigma_V}{\mu_V}$, where $\sigma_V$ and $\mu_V$ are the standard deviation, and the mean of the Voronoi tessellation areas of the users, respectively. CoV=1 corresponds to the Poisson point process, while CoV$>$1 represents clustered distribution of the users. The probability of being a delay-sensitive or delay-tolerant user is $0.2$ and $0.8$, respectively. All results are averaged over 100 Monte Carlo simulations. The urban environment and the simulation parameters are provided in \tablename~\ref{table1} and \tablename~\ref{table2}, respectively. 

\begin{table}[t]
	\renewcommand{\arraystretch}{1.3}
	\centering
	\caption{Urban Environment Parameters \cite{7037248}}\label{table1}
	\resizebox{5cm}{!}{
		\begin{tabular}[t]{|c |c |} 
			\hline
			\scriptsize\textbf{Parameter} & \scriptsize\textbf{Value} \\  
			\hline
			$(a,b)$ & (9.61, 0.16)  \\ 
			\hline
			$(\mu_{\text{LoS}},k_{\text{LoS}},l_{\text{LoS}})$ & (1 dB, 10.39, 0.05) \\
			\hline
			$(\mu_{\text{NLoS}},k_{\text{NLoS}},l_{\text{NLoS}})$ & (20 dB, 29.6, 0.03) \\
			\hline
		\end{tabular}}
	\bigskip
	\centering
	\caption{Simulation Parameters}\label{table2}
	\resizebox{\columnwidth}{!}{
\begin{tabular}[t]{| c | c | c | c |} 
	\hline
	\scriptsize\textbf{Parameter} & \scriptsize\textbf{Value} & \scriptsize\textbf{Parameter} & \scriptsize\textbf{Value} \\ 
	\hline
	$f_c$ & 2 GHz  & Noise power spectral density & -174 dBm/Hz \\  
	\hline
	$P_0$ & 46 dBm & $P_j, \forall j \in \mathcal {J} \backslash 0$ & 36 dBm  \\
	\hline
	$h_{\text{max}}$ & 500 m & System Bandwidth & 10 MHz  \\
	\hline
	\end{tabular}}
\end{table}

\begin{figure}[t]
	\begin{center}
		\includegraphics[width=3in]{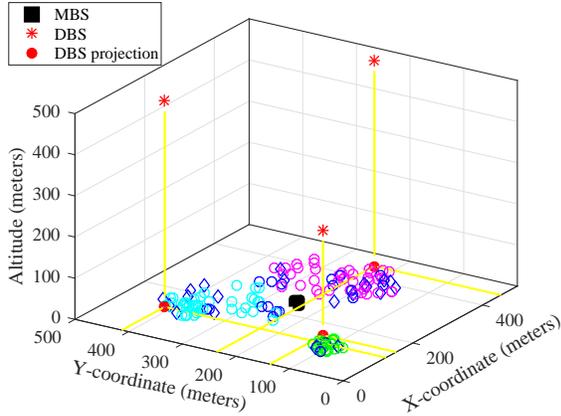}\\
	\end{center}
	\caption{A typical user distribution with CoV=3.3 along with the MBS and 3D placement of DBSs. The MBS is shown in a black square. DBSs and their projection on an XY-plane are shown using asterisk and red circles, respectively. Delay-sensitive and delay-tolerant users are shown by diamonds and circles, respectively. Also, different colours of users demonstrate association with different BSs.}
	\label{3D-dif-radius}
\end{figure}

A random realization of the user distribution along with the BSs are shown in Fig. \ref{3D-dif-radius}. The users that associate with different BSs are specified by different colours. It is observed that if a DBS has to serve farther users, it has to increase its altitude. Note that, a higher altitude creates a trade-off by yielding a larger probability of LoS links, as well as a higher path loss, as can be seen in \eqref{fspl} and \eqref{LoS}.

\begin{figure}[t]
	\begin{center}
		\includegraphics[width=3in]{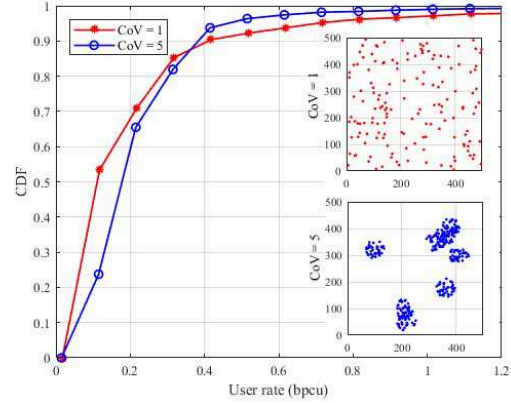}\\
	\end{center}
	\caption{CDF of users' rates for two different CoVs.}
	\label{cdf}
\end{figure}

Fig.~\ref{cdf} illustrates the empirical cumulative distribution functions (CDFs) of the users' rates for two different CoVs along with exemplary distributions corresponding to each CoV. It is observed that in a more clustered distribution, the probability that each user receives a higher rate increases. This confirms that the proposed algorithm can increase the performance of the cellular network in terms of users' satisfactions in more clustered distributions.

\begin{figure}[t]
	\begin{center}
		\includegraphics[width=3in]{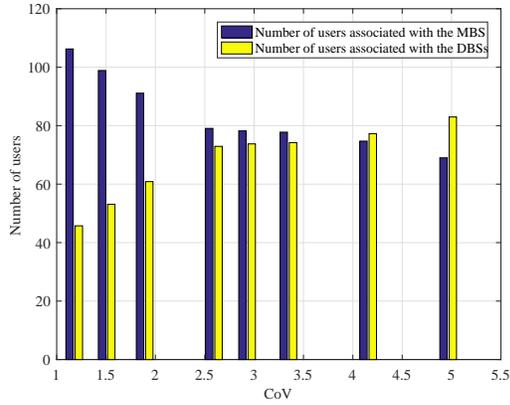}\\
	\end{center}
	\caption{Number of users associated with the MBS and the DBSs for different CoV values.}
	\label{num}
\end{figure}

The number of users associated with both the MBS and the DBSs are depicted in Fig. \ref{num}. By increasing the CoV, more users could be associated with the DBSs which results in better load balancing in the system. On the one hand, clustered users can be covered with DBSs at lower altitudes, which can increase SNR (and rate), similar to the case with green users in~Fig.~\ref{3D-dif-radius}. On the other hand, increasing the number of users served by a DBS, decreases the  wireless resources allocated to each user. Fig.~\ref{cdf} and Fig.~\ref{num} together show that this trade-off is in favor of rate, when CoV increases. 

\begin{figure}[h!]
	\begin{center}
		\includegraphics[width=3in]{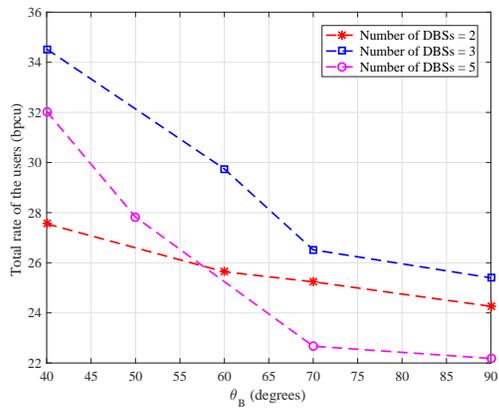}\\
	\end{center}
	\caption{Total rate of the users for different number of DBSs.}
	\label{theta}
\end{figure}

Finally, Fig. \ref{theta} shows the total rate of users associated with the DBSs for different half-power beamwidths, $\theta_B$s, and number of utilized DBSs. Note that, increasing $\theta_B$ increases the maximum possible coverage area. However, it also increases $D$ in \eqref{min-distance}, which means that to prevent overlapping, DBSs have to keep a larger distance between each other. Hence, in Fig.~\ref{theta}, the total capacity of users decreases, although the coverage radius increases with increasing $\theta_B$. Moreover, the effect of $\theta_B$ becomes more severe as the number of utilized DBSs increases. These results show that it is necessary to develop efficient interference cancellation methods for dense deployments of DBSs, since preventing overlaps between DBSs causes significant performance loss.   

\section{Conclusion}
In this study, delay-sensitive users are associated with the MBS, while delay-tolerant users can be associated with either one of the BSs. As all the DBSs share the same bandwidth, using directional antennas is proposed to relieve the effect of the interference. User-BS association and wireless backhaul bandwidth allocation are found through a decomposition method and the locations of DBSs are updated using a PSO algorithm. Also, further insights on the effects of CoV and half-power beamwidth is obtained by simulations. The results show that utilizing DBSs in cases where the users are clustered can increase total rate of the users associated with DBSs, despite depleting the resources. In order to prevent interference, overlaps of coverage areas of different DBSs are not allowed in many studies. However, the half-power beamwidth should be chosen carefully for these scenarios, as the results show that increasing the beamwidth can decrease total rate by preventing DBSs to be deployed in beneficial locations.   
%This paper provides a new DBS deployment plan considering user-BS associations, wireless backhaul bandwidth allocation, fairness, different user types, and 3D placement. Based on the various applications and different use cases in future wireless networks, users fall into two different categories: delay-sensitive (uRLLC users) and delay-tolerant (regular eMBB users).
\balance
\bibliographystyle{IEEEtran}
\bibliography{Xbib}

\end{document}